\documentclass[a4paper,12pt]{article}

\usepackage[utf8x]{inputenc}
\usepackage[T1]{fontenc}
\usepackage{times}
\usepackage[colorlinks=true,linkcolor=black,citecolor=black,urlcolor=blue]{hyperref}
\usepackage{geometry}
\usepackage{titlesec}
\titleformat{\section}
{\bfseries\uppercase}{\thesection.}{1em}{}
\titleformat{\subsection}
{\bfseries}{\thesection.\thesubsection.}{1em}{}

\usepackage{graphicx} % used to insert the figure
\usepackage{multirow} % used for the table
\usepackage[font=it]{caption}
\usepackage{cite}
\usepackage{breakurl}
\usepackage{indentfirst}
\usepackage{amsmath, amssymb, amsfonts, bm}
\usepackage{txfonts}
\usepackage{enumitem}
\usepackage{xcolor}

\titlespacing*{\subsection}{0pt}{1.5em}{0.2em}

\renewcommand\eqref[1]{Equation~\ref{#1}}

\renewcommand{\thesection}{\arabic{section}}
\renewcommand{\thesubsection}{\arabic{subsection}}

\begin{document}

\begin{center}
	\includegraphics[width=38.8mm, height=20.6mm]{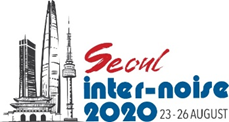}
\end{center}
\vskip.5cm

\begin{flushleft}
\fontsize{16}{20}\selectfont\bfseries
%\color{red}(The title should be written in "Times New Roman", 16point, bold font. The first alphabet of the first word in the title should be capitalized)  \\
\color{black}Active Noise Control based on the Momentum Multichannel Normalized Filtered-x Least Mean Square Algorithm 
\end{flushleft}
\vskip1cm

\renewcommand\baselinestretch{1}
\begin{flushleft}

Dongyuan Shi\footnote{dongyuan.shi@ntu.edu.sg}, Woon-Seng Gan\footnote{EWSGAN@ntu.edu.sg}, Bhan Lam\footnote{bhanlam@ntu.edu.sg}, Shulin Wen\footnote{alicia.wen@ntu.edu.sg}, and Xiaoyi Shen\footnote{XIAOYI003@e.ntu.edu.sg}\\
School of Electrical and Electronic Engineering, Nanyang Technological University.\\
50 Nanyang Ave, Singapore 639798.\\
%
%\vskip.5cm
%Given name Family Name2\footnote{mail2@example.com}\\
%Institution\\
%Full address\\

\end{flushleft}

\textbf{\centerline{ABSTRACT}\\
Multichannel active noise control (MCANC) is widely utilized to achieve significant noise cancellation area in the complicated acoustic field. Meanwhile, the filter-x least mean square (FxLMS) algorithm gradually becomes the benchmark solution for the implementation of MCANC due to its low computational complexity. However, its slow convergence speed more or less undermines the performance of dealing with quickly varying disturbances, such as piling noise. Furthermore, the noise power variation also deteriorates the robustness of the algorithm when it adopts the fixed step size. To solve these issues, we integrated the normalized multichannel FxLMS with the momentum method, which hence, effectively avoids the interference of the primary noise power and accelerates the convergence of the algorithm. To validate its effectiveness, we deployed this algorithm in a multichannel noise control window to control the real machine noise. 
}
\section{Introduction}
%The INTER-NOISE 2020 SEOUL Proceedings will be distributed to the congress participants on a memory stick.\par
%The purpose of these instructions is to ensure the uniformity of the publication.\par
%The manuscript should be submitted as a PDF file whose font is 12-point "Times New Roman". The length of a manuscript should be at most 12 pages and at least four pages.\par
%Only manuscripts in English will be accepted for the Proceedings.\par
%You must not insert any page number, header or foot note except the e-mail addresses in the first page of the manuscript~\cite{herranz19}.
Active noise control (ANC) is a technique that utilizes the loudspeaker generating "anti-noise" wave with the negative amplitude of the unwanted noise to cancel this acoustic disturbance~\cite{kajikawa_gan_kuo_2012,hansen2002understanding,Elliott1993Active,pawelczyk2008active}. Compared to passive noise cancellation strategy, such as the noise barrier, ANC exhibits more effectiveness in mitigating the low-frequency noise without occupying large space, affecting air ventilation, and destroying the natural environment. Hence, the ANC technique is widely applied in many different fields, including the headphones~\cite{chang2010active}, windows~\cite{he2019exploiting,LAM201816,MURAO2019338,hasegawa2018window,shi2017understanding,shi2017algorithms,shi2021block}, and the open space~\cite{Zhang2018Space}. However, in many real scenario, ANC can only achieve the local noise control around the error sensor~\cite{nelson1991active}. To enlarge the size of the noise reduction area, the multichannel ANC (MCANC) is usually applied at the expense of the system complexity~\cite{kuo1999active}.  

With the significant development of powerful processors and other digital devices~\cite{shi2016comparison}, such as digital signal processors, analog-to-digital converters (ADC), and digital-to-analog converters (DAC), it becomes feasible to realize the active control with adaptive algorithms~\cite{kuo2000review,Shi2020Practical}. Among these algorithms, the filtered-x least mean square (FxLMS) algorithm is prevalent in various applications since its low computational complexity~\cite{MorganMorganHistory}. Besides, it utilizes a filtered reference signal to update the control filter and compensates for the delay involved by the secondary path, which enhances the system robustness. Since its satisfactory performance in the single-channel ANC, FxLMS is also extended to the multichannel FxLMS (McFxLMS) algorithm while maintaining similar advantages in the MCANC application~\cite{elliott2000signal}.  Meanwhile, some other FxLMS-based algorithms have been proposed to further reduce computations~\cite{shi2017multiple,shi2018partial}, improve convergence~\cite{guo2020convergence,shi2019simulation}, or cope with the output saturation issue~\cite{qiu2001study,shi2017effect,shi2019practical,shi2019two,shi2019optimal,shi2021optimal}.   

Nevertheless, these FxLMS-based algorithms are always haunted by a practical issue~\cite{kuo1996active}: the step size bound is sensitive to the reference signal's power. An inappropriate step-size selection usually results in the divergence or slow convergence problem. For the same reason, the stability of the McFxLMS algorithm will be deteriorated in canceling the varying noise when its step size is chosen as a constant. 
To solve this issue, the variable step-size methods seem to be an effective strategy. However, most of these variable step-size mechanisms will severely aggravate the computational load, especially for the multichannel system. Under this situation, the multichannel normalized FxLMS (MNFxLMS) algorithm~\cite{chung2016multi,haykin1986adaptive} becomes a better choice because it can avoid the influence of input power by pre-whitening the referenced signal while slightly increasing computations. In that, MNFxLMS is undoubtedly suitable to deal with the primary noise with a massive power variation over time. To further improve its convergence, we integrate the momentum technique to MNFxLMS in this paper. In the new algorithm, the momentum term~\cite{ting2000tracking,Lok2004LMS,roy1990analysis} accumulates the previous gradient information to accelerate the convergence of MNFxLMS, which hence, leads to a satisfactory noise reduction performance when dealing with the quick-varying noise. The momentum MNFxLMS also smooths the varied gradient and reduces the high-frequency disturbance on the control filter's weight. Furthermore, this paper carries out the simulations on the proposed algorithm, which is used to deal with real quick-varying noise in measured paths.

This paper is organized as the following descriptions: Section~\ref{MNFxlMS_algorithm} revisits the multichannel normalized FxLMS algorithm; Section~\ref{Momentum} proposes the momentum MNFxLMS algorithm and addresses a brief analysis. Section~\ref{Simualtion} exhibits the simulation results of the McFxLMS, MNFxLMS, and momentum MNFxLMS algorithms, and Section~\ref{Conclusion} summaries the whole paper.

\section{The Multichannel Normalized Filtered-x LMS algorithm}
\label{MNFxlMS_algorithm}
\begin{figure}[!h]
	\centering
	\includegraphics[width=12cm]{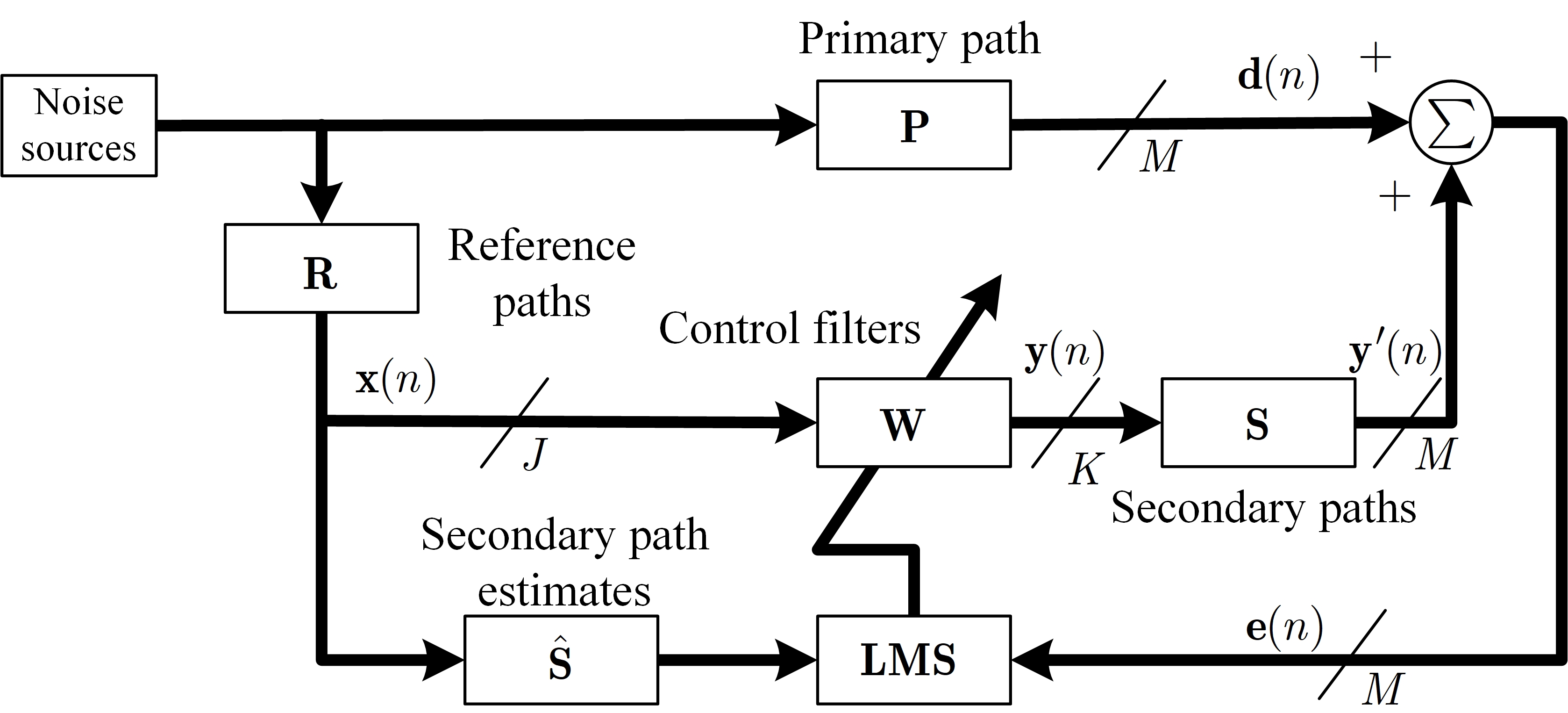}
	\caption{Block diagram of a multichannel ANC with $J$ microphones, $K$ secondary sources, and $M$ error microphones~\cite{IWAI2019151}.}
	\label{Fig_1_MCANC}
\end{figure}

In this paper, we consider a $J\times K\times M$ multichannel active noise control (MCANC) system, which uses $J$ reference microphones and $K$ secondary sources to cancel the disturbances at $M$ error microphones, as shown in Figure~\ref{Fig_1_MCANC}. In this figure, $\mathbf{P}$, $\mathbf{W}$, and $\mathbf{S}$ denote the transfer functions of the primary paths, control filters, and secondary paths, respectively. $\mathbf{\hat{S}}$ stands for the estimate of $\mathbf{S}$ and is obtained through the offline system identification.

The control signal of the $k$th secondary source can be expressed as 
%---------------------------Eq. 1---------------------------------------
\begin{equation}\label{eq_1}
	y_\mathrm{k}(n) = \sum^\mathrm{J}_\mathrm{j=1}\mathbf{w}^\mathrm{T}_\mathrm{kj}(n)\mathbf{x}_\mathrm{j}(n)
\end{equation}
%-----------------------------------------------------------------------  
where $\mathbf{w}_\mathrm{kj}(n)$ denotes the $kj$th control filter that models the $j$th reference to drive the $k$th secondary source and is expressed as 
%----------------------------------------------------------------------- 
\begin{equation}
	\mathbf{w}_\mathrm{kj}(n) = \begin{bmatrix}
	w_\mathrm{kj,1}(n) & w_\mathrm{kj,2}(n)	& \cdots & w_\mathrm{kj,N}(n)	
	\end{bmatrix}^\mathrm{T}\in\mathbb{R}^\mathrm{N\times 1}\notag
\end{equation}
%-----------------------------------------------------------------------
which has $N$ taps, and $\mathbf{x}_\mathrm{j}(n)$ is the $j$th reference vector give by
%----------------------------------------------------------------------- 
\begin{equation}
\mathbf{x}_\mathrm{j}(n) = \begin{bmatrix}
x_\mathrm{j}(n) & x_\mathrm{j}(n-1) & \cdots & x_\mathrm{j}(n-N-1)	
\end{bmatrix}^\mathrm{T}\in\mathbb{R}^\mathrm{N\times 1}.\notag
\end{equation}
%-----------------------------------------------------------------------
$\mathrm{T}$ and $\mathbb{R}$ represent the transpose operation and the real number, respectively. The error signal at the $m$th microphone can be written as 
%---------------------------Eq. 2---------------------------------------
\begin{equation}\label{eq_2}
	e_\mathrm{m}(n) = d_\mathrm{m}(n) + \sum^\mathrm{K}_\mathrm{k=1}y_\mathrm{k}(n)*s_\mathrm{mk}(n) 
\end{equation}
%---------------------------Eq. 2---------------------------------------
where $\ast$ denotes the linear convolution, and $s_\mathrm{mk}(n)$ stands for the secondary path from the $k$th secondary source to the $m$th error microphone.

Based on the principle of the minimal disturbance~\cite{haykin1986adaptive}, we define an increment of the $kj$th control filter at the $n+1$ iteration as   
%---------------------------Eq. 3---------------------------------------
\begin{equation}\label{eq_3}
	\delta\mathbf{w}_\mathrm{kj}(n+1) = \mathbf{w}_\mathrm{kj}(n+1) -\mathbf{w}_\mathrm{kj}(n)
\end{equation}
%---------------------------Eq. 3---------------------------------------
and an equality constrain as 
%---------------------------Eq. 4---------------------------------------
\begin{equation}\label{eq_4}
	d_\mathrm{m}(n) + \sum^\mathrm{J}_\mathrm{j=1}\sum^\mathrm{K}_\mathrm{k=1}\mathbf{w}^\mathrm{T}_\mathrm{kj}(n+1)\mathbf{x^\prime}_\mathrm{jkm}(n) = 0
\end{equation}
%---------------------------Eq. 4---------------------------------------
where $\mathbf{x^\prime}_\mathrm{jkm}(n)$ represents the $jkm$th filtered reference signal obtained from
%---------------------------Eq. 5---------------------------------------
\begin{equation}\label{eq_5}
	\mathbf{x^\prime}_\mathrm{jkm}(n) = \mathbf{x}_\mathrm{j}(n)*s_\mathrm{mk}(n)\in\mathbb{R}^\mathrm{N\times 1}.
\end{equation}
%---------------------------Eq. 5---------------------------------------

To solve \eqref{eq_3} and \eqref{eq_4}, we construct a Lagrange function as
%---------------------------Eq. 6---------------------------------------
\begin{equation}\label{eq_6}
	J(n) = \sum^\mathrm{J}_\mathrm{j=1}\sum^\mathrm{K}_\mathrm{k=1}\big\|\delta\mathbf{w}_\mathrm{kj}(n+1)\big\|^2 +\sum^\mathrm{M}_\mathrm{t=1}\lambda_\mathrm{t}\left[d_\mathrm{t}(n) + \sum^\mathrm{J}_\mathrm{j=1}\sum^\mathrm{K}_\mathrm{k=1}\mathbf{w}^\mathrm{T}_\mathrm{kj}(n+1)\mathbf{x^\prime}_\mathrm{jkt}(n) \right]
\end{equation}
%---------------------------Eq. 6---------------------------------------
where $\|\cdot\|$ and $\lambda_\mathrm{t}$ denote the 2-norm and the $t$th Lagrange multiplier, respectively. 

According to the Lagrange multiplier method~\cite{haykin1986adaptive}, we derive the solution to minimize \eqref{eq_6} as the following procedures: 

(1). The gradient of \eqref{eq_6} with respect to $\mathbf{w}_\mathrm{kj}(n+1)$ is derived as 
%---------------------------Eq. 7---------------------------------------
\begin{equation}\label{eq_7}
	\frac{\partial J(n)}{\partial \mathbf{w}_\mathrm{kj}(n+1)}=2\left[\mathbf{w}_\mathrm{kj}(n+1) -\mathbf{w}_\mathrm{kj}(n)\right]+\sum^\mathrm{M}_\mathrm{t=1}\lambda_\mathrm{t}\mathbf{x^\prime}_\mathrm{jkt}(n). 
\end{equation}
%---------------------------Eq. 7---------------------------------------
Setting \eqref{eq_7} to $\mathbf{0}$ yields
%---------------------------Eq. 8---------------------------------------
\begin{equation}\label{eq_8}
	\mathbf{w}_\mathrm{kj}(n+1) = \mathbf{w}_\mathrm{kj}(n) - \frac{1}{2}\sum^\mathrm{M}_\mathrm{t=1}\lambda_\mathrm{t}\mathbf{x^\prime}_\mathrm{jkt}(n)
\end{equation}
%---------------------------Eq. 8---------------------------------------

(2). By substituting \eqref{eq_8} into \eqref{eq_4}: 
%---------------------------Eq. 9---------------------------------------
\begin{equation}\label{eq_9}
		d_\mathrm{m}(n) = -\sum^\mathrm{J}_\mathrm{j=1}\sum^\mathrm{K}_\mathrm{k=1}\left[\mathbf{w}_\mathrm{kj}(n) - \frac{1}{2}\sum^\mathrm{M}_\mathrm{t=1}\lambda_\mathrm{t}\mathbf{x^\prime}_\mathrm{jkt}(n)\right]^\mathrm{T}\mathbf{x^\prime}_\mathrm{jkm}(n)
\end{equation}
%---------------------------Eq. 9---------------------------------------
we can obtain 
%---------------------------Eq.10---------------------------------------
\begin{equation}\label{eq_10}
	e_\mathrm{m}(n)=\frac{1}{2}\sum^\mathrm{J}_\mathrm{j=1}\sum^\mathrm{K}_\mathrm{k=1}\sum^\mathrm{M}_\mathrm{t=1}\lambda_\mathrm{t}\mathbf{x^\prime}^\mathrm{T}_\mathrm{jkt}(n)\mathbf{x^\prime}_\mathrm{jkm}(n)\approx\frac{1}{2}\sum^\mathrm{J}_\mathrm{j=1}\sum^\mathrm{K}_\mathrm{k=1}\lambda_\mathrm{m}\big\|\mathbf{x^\prime}_\mathrm{jkm}(n)\big\|^2
\end{equation}
%---------------------------Eq.10---------------------------------------
where it is assumed that $\mathbf{x^\prime}_\mathrm{jkt}(n)$ and $\mathbf{x^\prime}_\mathrm{jkm}(n)$ are orthogonal ($t\ne m$)~\cite{chung2016multi}. Hence, from \eqref{eq_10}, the Lagrange multiplier is derived as 
%---------------------------Eq.11---------------------------------------
\begin{equation}\label{eq_11}
	\lambda_\mathrm{m} = \frac{2 e_\mathrm{m}(n)}{\sum^\mathrm{J}_\mathrm{j=1}\sum^\mathrm{K}_\mathrm{k=1}\big\|\mathbf{x^\prime}_\mathrm{jkm}(n)\big\|^2}.
\end{equation}
%---------------------------Eq.11--------------------------------------- 

(3). Substituting \eqref{eq_11} into \eqref{eq_8} yields 
%---------------------------Eq.12--------------------------------------- 
\begin{equation}\label{eq_12}
\mathbf{w}_\mathrm{kj}(n+1) = \mathbf{w}_\mathrm{kj}(n) - \sum^\mathrm{M}_\mathrm{m=1}\frac{e_\mathrm{m}(n)}{\sum^\mathrm{J}_\mathrm{j=1}\sum^\mathrm{K}_\mathrm{k=1}\big\|\mathbf{x^\prime}_\mathrm{jkm}(n)\big\|^2}\mathbf{x^\prime}_\mathrm{jkm}(n).
\end{equation}
%---------------------------Eq.12--------------------------------------- 
To control the magnitude of the increment of the control filter, we introduce a positive multiplier $\widetilde{\mu}~(0<\widetilde{\mu}<1)$ in \eqref{eq_12}:
%---------------------------Eq.13--------------------------------------- 
\begin{equation}\label{eq_13}
\mathbf{w}_\mathrm{kj}(n+1) = \mathbf{w}_\mathrm{kj}(n) - \widetilde{\mu}\sum^\mathrm{M}_\mathrm{m=1}\frac{e_\mathrm{m}(n)\mathbf{x^\prime}_\mathrm{jkm}(n)}{\sum^\mathrm{J}_\mathrm{j=1}\sum^\mathrm{K}_\mathrm{k=1}\big\|\mathbf{x^\prime}_\mathrm{jkm}(n)\big\|^2+\varepsilon}
\end{equation}
%---------------------------Eq.13---------------------------------------
where $\varepsilon$ is a small positive scalar to guarantee the division result within the finite value. \eqref{eq_13} is so-call multichannel normalized filtered-x least mean square (MNFxLMS) algorithm~\cite{chung2016multi}. \eqref{eq_13} can be rewritten as 
%-----------------------------------------------------------------------
\begin{equation}
\mathbf{w}_\mathrm{kj}(n+1) = \mathbf{w}_\mathrm{kj}(n) - \sum^\mathrm{M}_\mathrm{m=1} \mu_\mathrm{m}(n)e_\mathrm{m}(n)\mathbf{x^\prime}_\mathrm{jkm}(n)
\end{equation}
%-----------------------------------------------------------------------
where the equivalent step size of the MNFxLMS algorithm is given by 
%---------------------------Eq.14---------------------------------------
\begin{equation}\label{eq_14}
	\mu_\mathrm{m} (n) = \frac{\widetilde{\mu}}{\sum^\mathrm{J}_\mathrm{j=1}\sum^\mathrm{K}_\mathrm{k=1}\big\|\mathbf{x^\prime}_\mathrm{jkm}(n)\big\|^2+\varepsilon}
\end{equation}
%---------------------------Eq.14--------------------------------------- 
which is inversely proportional to the power of the reference signal. Therefore, the MNFxLMS algorithm effectively avoids the influence of input power variation and enforces the adaptive algorithm's robustness.  

\section{The Momentum MNFxLMS algorithm}  
\label{Momentum}
To further fasten the convergence of the MNFxLMS algorithm, we integrate the momentum mechanism~\cite{ting2000tracking,Lok2004LMS,roy1990analysis} into the updating equation of \eqref{eq_13} as 
%---------------------------Eq.15---------------------------------------
\begin{equation}\label{eq_15}
	\mathbf{w}_\mathrm{kj}(n+1) = \mathbf{w}_\mathrm{kj}(n) -\eta_\mathrm{kj}(n)
\end{equation}
%---------------------------Eq.15---------------------------------------  
where $\eta_\mathrm{kj}(n)$ denotes the momentum of the algorithm given by 
%---------------------------Eq.16---------------------------------------  
\begin{equation}\label{eq_16}
	\eta_\mathrm{jk}(n) =\gamma\cdot\eta_\mathrm{kj}(n-1) + \widetilde{\mu}\sum^\mathrm{M}_\mathrm{m=1}\frac{e_\mathrm{m}(n)\mathbf{x^\prime}_\mathrm{jkm}(n)}{\sum^\mathrm{J}_\mathrm{j=1}\sum^\mathrm{K}_\mathrm{k=1}\big\|\mathbf{x^\prime}_\mathrm{jkm}(n)\big\|^2+\varepsilon}.
\end{equation}
%---------------------------Eq.16--------------------------------------- 
In \eqref{eq_16}, $\gamma \in \left(0,1\right)$ stands for the forgetting factor, which decides the degree of the influence of previous gradients on the weight increment. Since the momentum term of \eqref{eq_16} accumulates the previous gradients, it is evident that the convergence of the proposed algorithm will be significantly accelerated if these gradients have the same direction.

\begin{figure}[!h]
	\centering
	\includegraphics[width=12cm]{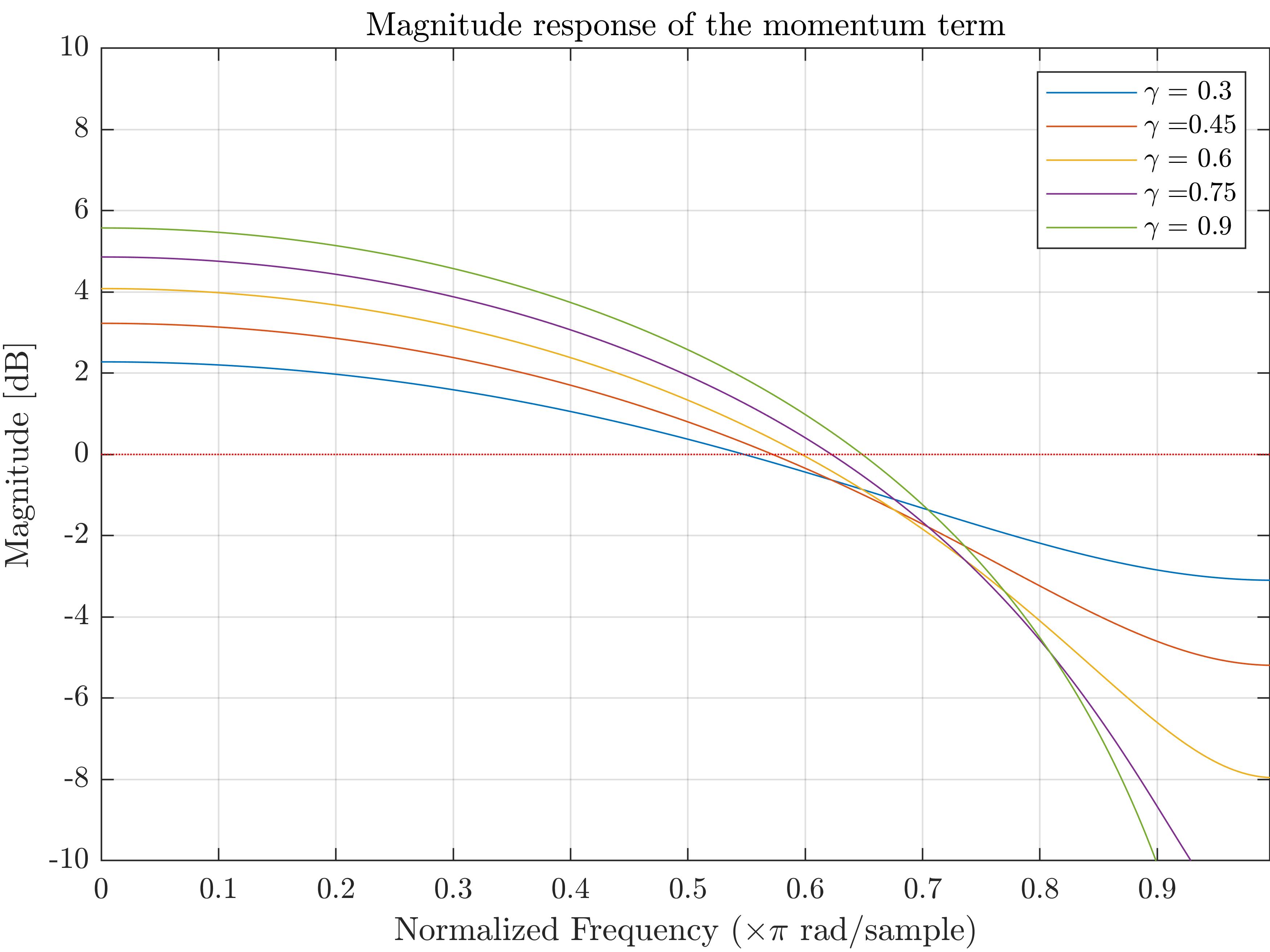}
	\caption{Magnitude response of the momentum function with different values of the forgetting factor $\gamma$.}
	\label{Fig_Momemtum}
\end{figure}

It is worth noting that the z-transform expression of \eqref{eq_16} can be written as  
%---------------------------Eq.16---------------------------------------  
\begin{equation}\label{eq_17}
	H(z) = \frac{1}{1-\gamma z^{-1}} \Delta(z)
\end{equation}
%---------------------------Eq.16---------------------------------------  
where $H(z)$ and $\Delta (z)$ represent the z-transform of $\eta (n)$ and the last term in the left side of \eqref{eq_16}, respectively. The magnitude response of \eqref{eq_17} is shown in Figure~\ref{Fig_Momemtum}. It can figure out that, for the quick-varying gradient, the momentum term works like a low-pass filter, which attenuates the high-frequency disturbance on the control filter's weights. However, for the low-frequency varied gradient, its amplitude will be amplified so that to improve the convergence of the algorithm.    

\section{Simulation result} 
\label{Simualtion}
\begin{figure}[!h]
	\centering
	\includegraphics[width=12cm]{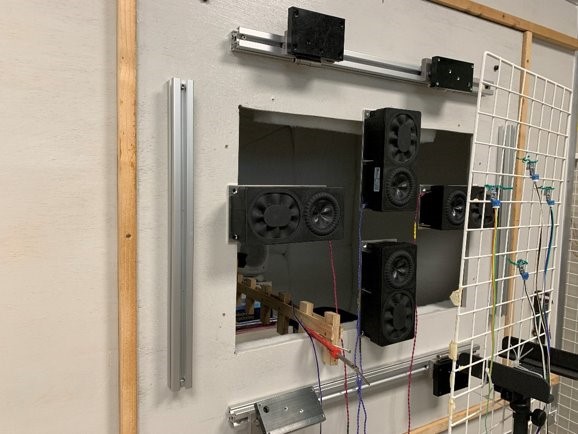}
	\caption{The front view of a $4$-channel active noise control system, which has $1$ reference microphone, $4$ secondary source and $4$ error sensors.}
	\label{Fig_platform}
\end{figure} 
\begin{figure}[!h]
	\centering
	\includegraphics[width=12cm]{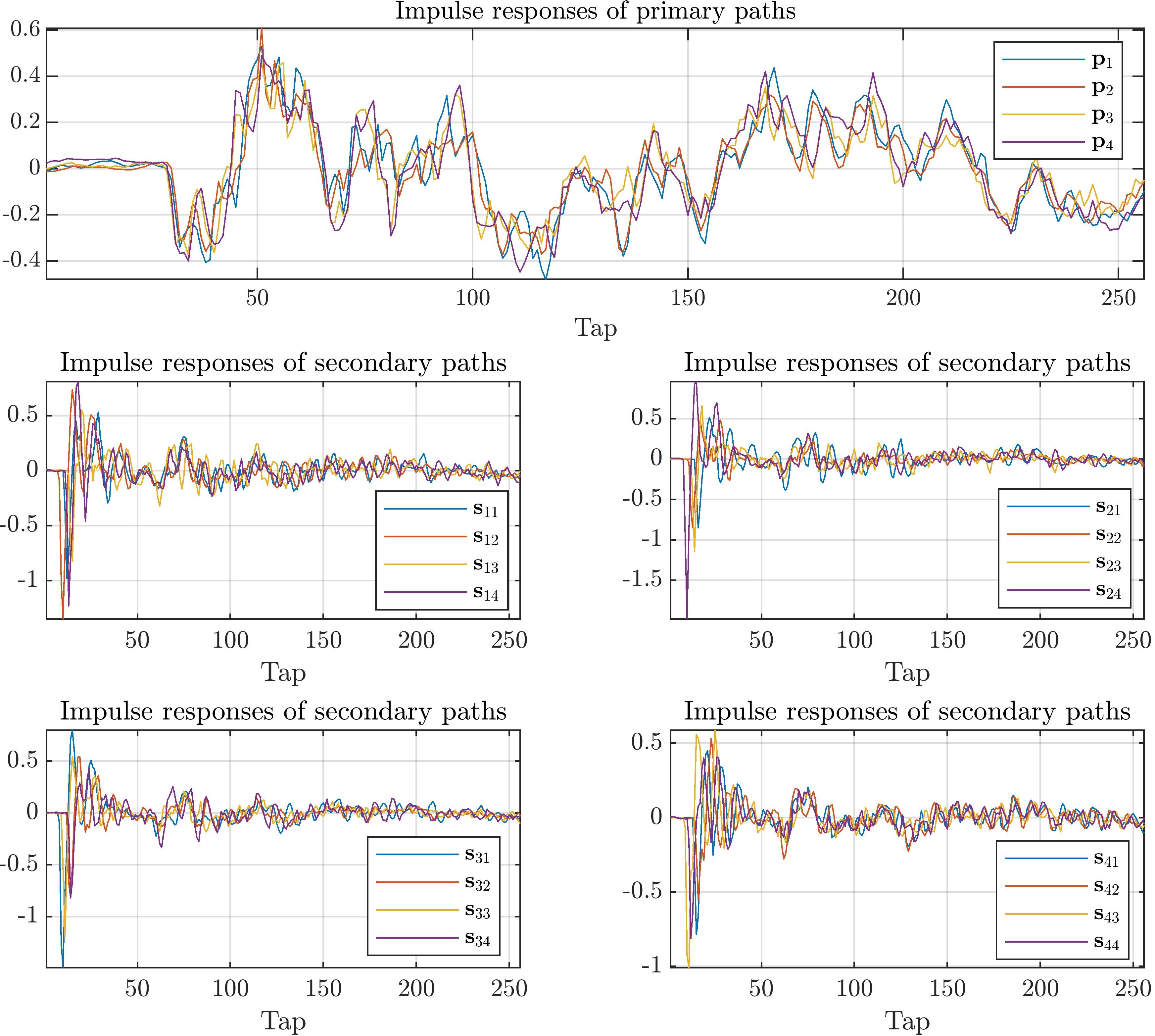}
	\caption{Impulse responses of primary paths and secondary paths: $\mathbf{p}_\mathrm{j}$ denotes the path from the primary source to the $j$th error microphone; $\mathbf{s}_\mathrm{mk}$ represents the secondary path form the $k$th secondary source to the $m$th error microphone, and $j,k,m = 1,2,3,4$.}
	\label{Fig_2_Paths}
\end{figure} 
To carry out the simulations on the McFxLMS, MNFxLMS, and momentum MNFxLMS algorithms, we measured the primary and secondary paths from a 4-channel ANC system installed in a noise chamber, as shown in Figure~\ref{Fig_platform}. This wooden chamber has a dimension of $1.2~\text{m}\times 1.2~\text{m} \times 1.2~\text{m}$ and a aperture with the size of $60~\text{cm}\times 50~\text{cm}$ at its facade. A noise source is placed inside the chamber and put $1 \text{m}$ way from the aperture, which has four secondary mounted around its frame. There are four error microphones fixed in a grid kept $50~\text{cm}$ away from the secondary sources. The reference microphone of this MCANC system is placed close to the noise source. The impulse responses of measured primary and secondary paths are illustrated in Figure~\ref{Fig_2_Paths}.

Furthermore, the control filter and secondary path estimate in all algorithms have $512$ and $256$ taps, respectively. The forgetting factor of the momentum MNFxLMS algorithm is set to $0.9$.  

\subsection{The cancellation of varying broadband noise}
\begin{figure}[!h]
	\centering
	\includegraphics[width=12cm]{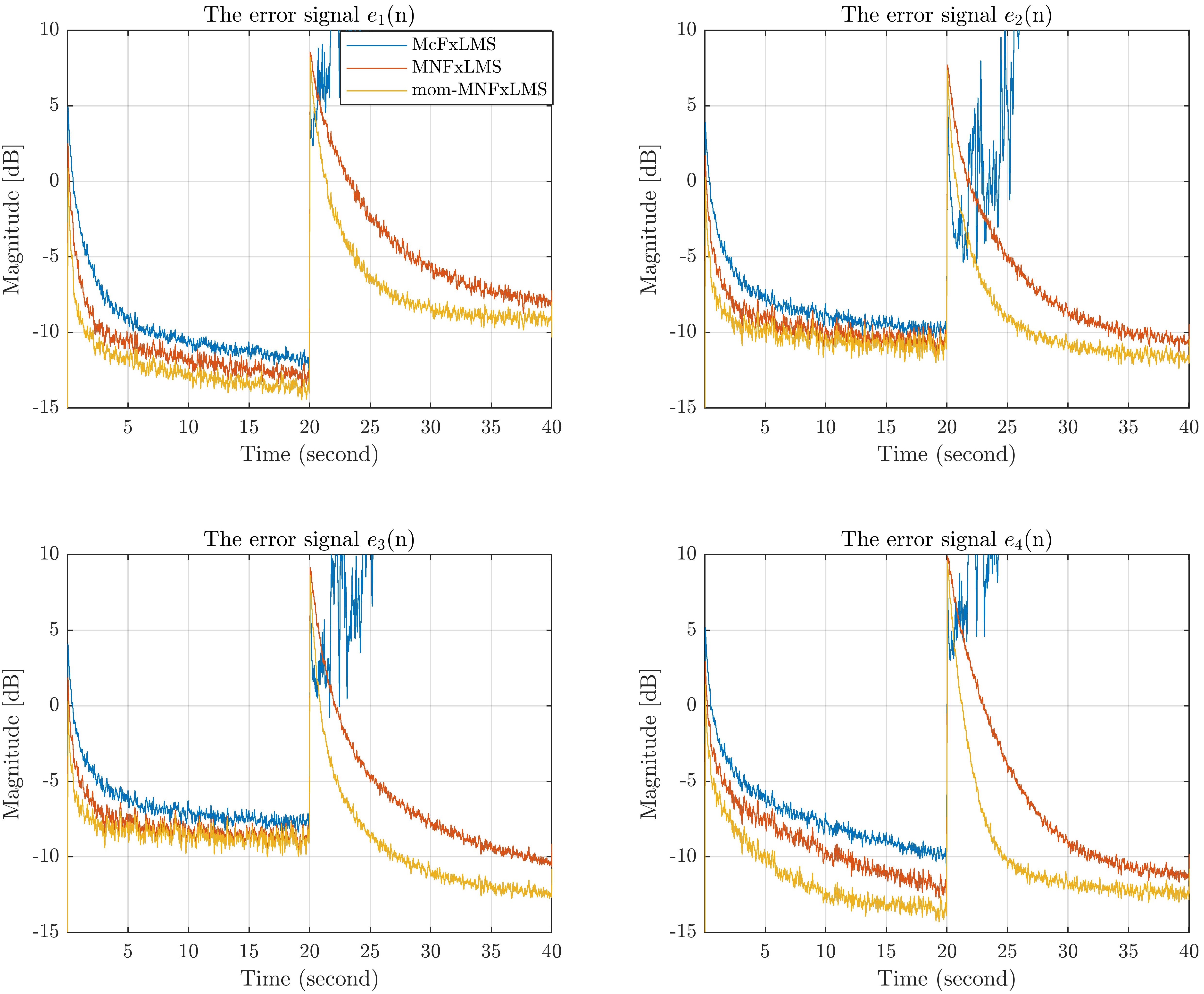}
	\caption{Time histories of error signals at the four error microphones. The step size of the McFxLMS algorithm is set to $0.000001$, and step sizes of the MNFxLMS and momentum MNFxLMS algorithms are set to $0.001$. The forgetting factor of $\gamma$ is $0.9$.}
	\label{Fig_Varied_Noise}
\end{figure} 

In the first simulation, the 4-channel ANC system is utilized to cancel the broadband noise. In the first $20$ seconds, the broadband noise has the frequency of $200$ to $800$ Hz and the amplitude of $10$ dB; in the next $20$ seconds, the amplitude of this noise changes to $15$ dB, and its frequency becomes $100$ to $1600$ Hz.  

During the simulation, the McFxLMS algorithm has a step size of $0.000001$, while the step sizes of the MNFxLMS and momentum MNFxLMS are set to $0.001$. Figure~\ref{Fig_Varied_Noise} exhibits the four error signal's waveform of the algorithms. From this result, it can be found that the McFxLMS algorithm diverges when it meets the large input power. Meanwhile, the MNFxLMS and momentum MNFxLMS algorithm can remain a satisfactory converge even though the primary noise has a significant variation. It is because the step size in MNFxLMS will automatically adjust with the input power, as shown in \eqref{eq_14}. Furthermore, with the assistance of the gradient accumulation, the momentum MNFxLMS algorithm shows a faster convergence behavior than the conventional MNFxLMS.  

\subsection{The cancellation of a real piling noise}
\begin{figure}[!h]
	\centering
	\includegraphics[width=12cm]{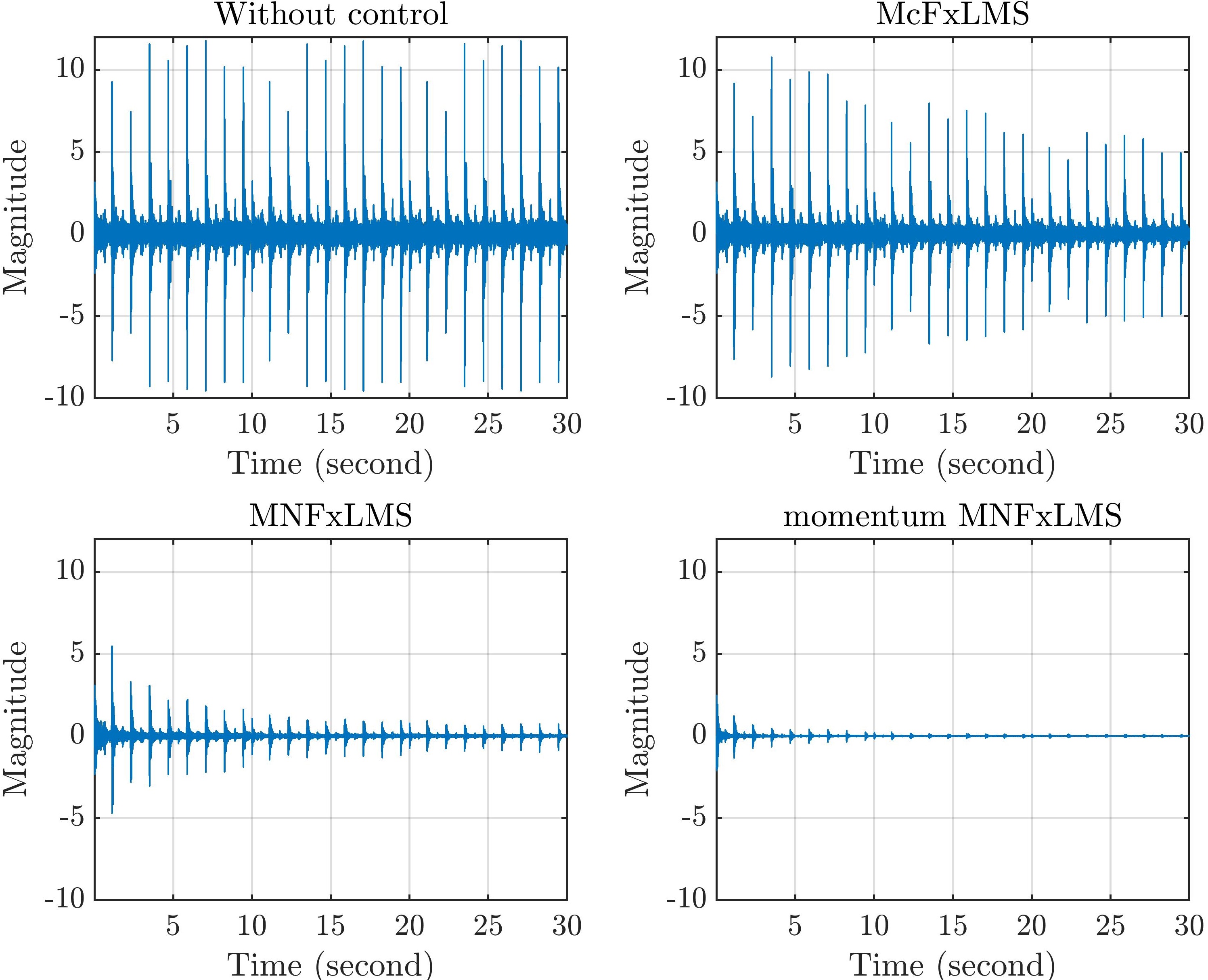}
	\caption{Time history of a real piling noise at the first error microphone. The step size of the McFxLMS algorithm is set to $0.0001$, and the step sizes of the MNFxLMS and momentum MNFxLMS algorithms are set to $0.001$. The forgetting factor of $\gamma$ is $0.9$.}
	\label{Fig_RealTrack}
\end{figure} 

In the second simulation, the primary noise becomes a real piling noise, as shown in Figure~\ref{Fig_RealTrack}. To ensure the convergence of the McFxLMS algorithm, its step size is set to $0.0001$ at the expense of its convergence speed. Meanwhile, the MNFxLMS and momentum MNFxLMS can have a greater step size of $0.001$. Under this situation, the momentum MNFxLMS algorithm achieves the fastest convergence among these algorithms, as illustrated in Figure.~\ref{Fig_RealTrack}. However, all these three algorithms obtain similar noise reduction levels around $20$ dB at the steady-state.

\subsection{The cancellation of an fMRI noise}
\begin{figure}[!h]
	\centering
	\includegraphics[width=12cm]{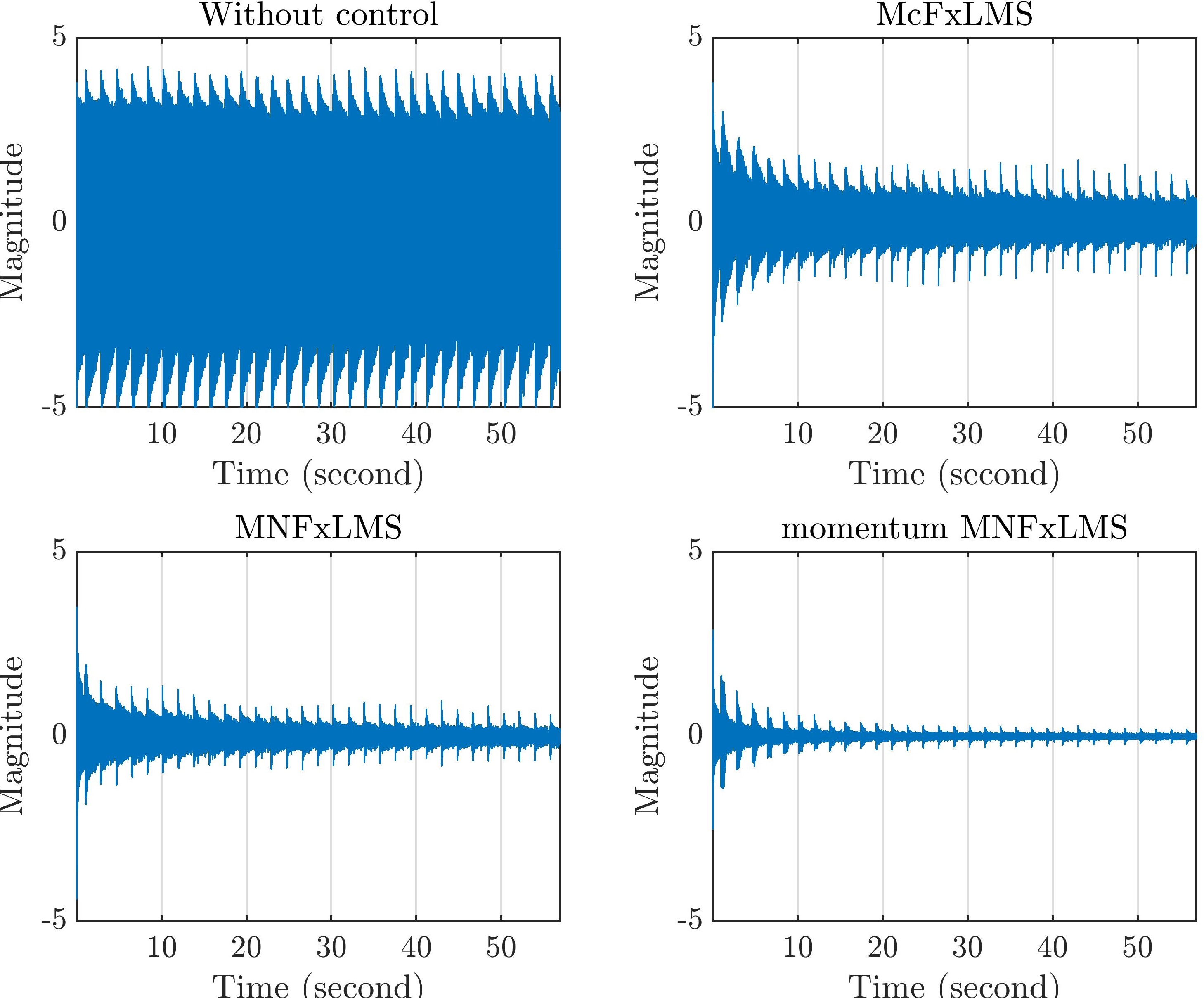}
	\caption{Time history of an fMRI noise at the first error microphone. The step size of the McFxLMS algorithm is set to $0.0001$, and the step sizes of the MNFxLMS and momentum MNFxLMS algorithms are set to $0.01$. The forgetting factor of $\gamma$ is $0.9$.}
	\label{Fig_RealTrack_fmri}
\end{figure}        

In the final simulation, the primary noise is a real functional magnetic resonance imaging (fMRI) machine noise~\cite{miyazaki2012head}, as shown in Figure~\ref{Fig_RealTrack}. To ensure the convergence of the McFxLMS algorithm, its step size is set to $0.0001$. The MNFxLMS and momentum MNFxLMS algorithms have a step size of $0.001$. Under this situation, the momentum MNFxLMS algorithm achieves the fastest convergence among these algorithms, as illustrated in Figure.~\ref{Fig_RealTrack}. All these three algorithms obtain similar noise reduction levels around $21.2$ dB at the steady-state.

\section{Conclusions}
\label{Conclusion}
To assist the multichannel active noise control (MCANC) system in canceling quick-varying noise, the paper integrates the momentum method with the multichannel normalized filtered-x least mean square (MNFxLMS) algorithm. The MNFxLMS algorithm avoids the input power's influence on the step size bound, and the momentum method accelerates the convergence by accumulating the previous gradient information. The momentum MNFxLMS algorithm combines these two advantages and hence, exhibits a satisfactory performance in canceling the quick-varying noise. The simulation on the measured paths of a 4-channel ANC system verifies the effectiveness of the proposed algorithm in dealing with the real piling and fMRI noise. 

\section{Acknowledgements}
%We  gratefully acknowledge the authors for submitting their work to INTER-NOISE 2020 SEOUL.
This research is supported by the Singapore Ministry of National Development and the National Research Foundation, Prime Minister’s Office under the Cities of Tomorrow (CoT) Research Programme (CoT Award No. COT-V4-2019-1). Any opinions, findings, and conclusions or recommendations expressed in this material are those of the author(s) and do not reflect the views of the Singapore Ministry of National Development and National Research Foundation, Prime Minister’s Office, Singapore.

\bibliography{biblio} 
\bibliographystyle{ieeetr}

\end{document}